\documentstyle[12pt]{article}
\begin{document}
\title{{\bf Return of the Boltzmann Brains}
\thanks{Alberta-Thy-14-06, hep-th/0611158}}
\author{
Don N. Page
\thanks{Internet address:
don@phys.ualberta.ca}
\\
Institute for Theoretical Physics\\
Department of Physics, University of Alberta\\
Room 238 CEB, 11322 -- 89 Avenue\\
Edmonton, Alberta, Canada T6G 2G7
}
\date{(2006 November 15)}

\maketitle
\large
\begin{abstract}
\baselineskip 14.5 pt

	Linde in hep-th/0611043 shows that some (though not all)
versions of the global (volume-weighted) description avoid the
``Boltzmann brain'' problem raised in hep-th/0610079 if the universe
does not have a decay time less than 20 Gyr.  Here I give an apparently
natural version of the volume-weighted description in which the problem
persists, highlighting the ambiguity of taking the ratios of infinite
volumes that appear to arise from eternal inflation.

\end{abstract}
\normalsize

	One of the most important challenges in theoretical cosmology
today is the measure problem, the problem of giving theoretical
predictions of the relative measures or probabilities for various types
of observations.  A quantum solution of this problem might be a
specification of a quantum state of the universe that gives
normalizable expectation values of positive operators associated with
observations.  (In \cite{SQM,MS,Page-in-Carr} I have suggested that the
fundamental observations are conscious perceptions, each set of which
has its own positive quantum `awareness operator' whose expectation
value in the quantum state of the cosmos gives the measure of the
corresponding set of conscious perceptions, but other theorists might
prefer to interpret observations differently.)

	A severe problem has arisen from the predictions of eternal
inflation
\cite{Let1,Vet1,Set,Let2,Let3,Let4,Let5,Let6,Let7,Vet2,Get,Vet3}, which
seems to lead to an arbitrarily large universe, with an arbitrarily
large number of observers, which makes it problematic how to calculate
the probability of various observed features by taking the ratio of the
numbers or measures of the corresponding infinite sets of observers
\cite{LM,LLM,GLL,GL,Vil,WV,VVW,GV,GSVW,ELM,Teg,FHW,Lyth,LMuk,HWY,
Bousso}.  This problem is highlighted by paradoxes, if the universe
lasts too long, of too many disordered observations being made by
brains resulting from thermal fluctuations \cite{DKS,GKS,Sus03}
(``Boltzmann brains'' or BB's) or from vacuum fluctuations
\cite{lifetime,decay,HH-challenge} (``brief brains'' or bb's, though I
shall often lump them all together as BB's or BBs, not to be confused
with the BBS or Blum-Blum-Shub pseudorandom number generator
\cite{BBS}).  For example, it was found \cite{decay} that if the
universe continues to expand exponentially at the rate given by present
observations, then per comoving volume there would be an infinite
expectation value of the 4-volume if the expected decay time is less
than 13--19 Gyr (say 20 Gyr to give a conservative upper limit),
leading to very tiny probabilities of observations as ordered as ours
are.

	Bousso and Freivogel \cite{BF} have recently argued that this
paradox does not arise in the local description \cite{local} (unless
the decay time is much, much longer) and is evidence against the global
description of the multiverse.  On the other hand, Linde \cite{LindeBB}
has more recently given an example of how one may avoid the dreaded
``invasion of Boltzmann brains'' within his `standard' volume-weighted
distribution in the global description.  Here I wish to point out a
natural different regularization of the volume-weighted distribution in
which the problem of the Boltzmann brains returns. 

	Linde's solution in Section 8.3 of \cite{LindeBB} has at least
two de Sitter vacuum states, one (labeled by the subscript 1) with a
small expansion rate $H_1$ consistent with our observations, and
another (labeled by the subscript 2) with a larger expansion rate
$H_2$.  Then as a function of the global cosmic proper time $t$,
spacetime in state 2 will expand rapidly (asymptotically giving the
main contribution to the volume growth of both types of regions) and
will with some transition rate $\Gamma_{21}$ convert to state 1 where
both ordinary observers (OOs, ``honest folk like ourselves''
\cite{LindeBB}) and Boltzmann or brief brains (BBs) can exist.  Then
even though each spacetime region in state 1 will expand indefinitely
and produce far more BBs than OOs, at each time more spacetime region
in state 1 is being produced by transitions from the more rapidly
expanding spacetime in state 2, so that asymptotically at each large
but fixed proper time, nearly all of the spacetime regions in state 1
will arise from recent transitions from state 2, rather than from the
slower expansion of state 1.  Thus at each time, most of the volume in
state 1 will be early in its evolution, where OOs rather than BBs
dominate.  In particular, if $3(H_2-H_1) \gg \Gamma_{1B}$, the rate of
the BB production in state 1, then at each fixed late proper time,
there will be far more ordinary observers than Boltzmann brains or
brief brains from vacuum fluctuations.

	For example, in \cite{HH-challenge} I made the estimate that
the probability per 4-volume for a brief brain is $\Gamma_{1B}\sim
e^{-10^{42}}$, where the upper exponent had already been predicted
$3^3$ years ago \cite{42}.  This rate is indeed much, much less than
any reasonable $3(H_2-H_1)$.

	Linde's solution is analogous to the following situation. 
Consider imaginary humans who have a `youthful' phase of 100 years of
life with frequent and mostly ordered observations, followed by a
`senile' phase of trillions of years of infrequent and mostly
disordered observations.  Assume that the trillions of years are
sufficient to give more many total `senile' disordered observations
than `youthful' ordered observations for each human.  One might think
that most observations would then be disordered, so that someone's
having an ordered observation (which would thus be very improbable
under this scenario) could count as evidence against the theory giving
this scenario.  However, if the population growth rate of such humans
is sufficiently high that at each time the number of youthful humans
and their ordered observations outnumbers the senile humans and their
disordered observations, Linde's solution is that at each time the
probability is higher that an observation would be ordered than that it
would be disordered.

	I agree that Linde's solution is a possible way to regulate the
infinity of observations that occur in a universe that expands forever,
but it is not the only way, or, I might suggest, the most natural way. 
The problem arises from the fact that if the youthful humans or OOs are
always at late times to outnumber the senile humans or BBs, the
population of these fictitious humans, or the volume of the universe in
the original example, must continue growing forever, producing an
infinite number of both youthful and senile fictitious humans or of
both OOs and BBs in cosmology.  Then it is ambiguous how one takes the
ratio, which is the fundamental problem with trying to solve the
measure problem in theories with eternal inflation.  Linde himself, in
Section 8.2 of \cite{LindeBB}, gives a ``pseudo-comoving volume-weighted
distribution'' which he shows does have the problem of BBs dominating if
the decay rate of state 1 is not higher than the BB production rate. 

	For the analogue with the fictitious humans, an alternate way
to regulate the infinities (which seems simpler and more natural to me)
would be just to count the observations of each such human, regardless
of when they occur.  Then under the given scenario, the senile ones
would dominate, giving a low probability for having an ordered
observation.

	Similarly, in cosmology with spacetime regions 1 which allow
observations and spacetime regions 2 which for simplicity I shall
assume do not allow observations, an alternate simple way to regulate
the infinities would be just to take the observers in a single
connected region of spacetime in state 1.  If this region was born by a
transition from state 2 and died with a transition back to state 2 (or
to any other state), then since by hypothesis observers can only exist
within regions 1, it would be most natural to regulate the infinity
that arises from the infinite number of regions 1 that occur by
calculating the ratios of measures for observations within a single
such connected region 1.   Then if regions 1 with asymptotically
exponential growth rates $H_1$ agreeing with observations in our part
of the universe have half-lives longer than about 20 billion years,
nearly all observations made within it would be by BBs, which would
give almost entirely disordered observations, so our ordered
observations would be very unlikely according to this theory, counting
as strong observational evidence against it.

	One might still worry that even our connected region that is
entirely in state 1 might be spatially infinite (e.g., if $k=0$ or
$k=-1$), giving an infinite number of both OOs and BBs.  In this case
one would indeed need some other regularization in addition to that
above (focusing on only one connected region), such as using only a
finite comoving volume.  However, if our region came from inflation in
a finite space, such as a $k=+1$ universe, then we could still use the
entire spatial volume, which at each finite time would be finite.  If
the universe does expand faster than its decay rate (e.g., if its
half-life is greater than 20 Gyr), then the total future expectation
value of its 4-volume would be infinite, leading to an infinite
expected number of BBs, but there would only be a finite number of OOs,
so one could still conclude that the relative probability of an OO was
zero, and of the BBs, only a very tiny probability would be of ordered
observations.  Therefore, our ordered observations would be strong
evidence against this theory with this natural way of doing the
regularization (focusing on just one connected region 1).

	Thus Linde's 'standard' regularization in Section 8.3 of
\cite{LindeBB} is only one way to get finite ratios from the infinitely
many observers that appear to arise from eternal inflation, and besides
Linde's alternative prescription in Section 8.2 of \cite{LindeBB}, I
have shown above that an apparently natural way, still using volume
weighting, returns the problem of the Boltzmann brains.

	Of course, the fact that we have ordered observations and are
almost certainly not Boltzmann brains is strong evidence against what I
have here proposed as a natural way of using volume weighting in the
global viewpoint (unless the universe really is decaying with a
half-life less than 20 billion years \cite{decay}, which also seems
rather implausible \cite{decay,BF,LindeBB}).  So in comparison with the
observations, and under the assumption that the universe is not
decaying within 20 billion years, my proposal is definitely worse than
Linde's `standard' prescription.

	I mainly wish to highlight my opinion that the `standard'
prescription, with its emphasis on the global cosmic time at the end of
the rapid inflationary period at the beginning of our region 1, is not
an unambiguously natural proposal.  I am happy to praise it as a clever
step forward toward our understanding of the severe measure problem in
cosmology (whose solution must be part of our ultimate theory, as Linde
has emphasized in Section 9 of \cite{LindeBB}).  However, I do not yet
feel that it is sufficiently unique or natural to be {\it the} final
solution.  So the purpose of the present paper is not to be critical of
the valiant attempts to solve the measure problem, or to propose a
solution that I think is better, but to highlight my opinion that the
problem persists and to stimulate more intelligent OOs to look for a
natural solution that I myself don't begin to see how to solve.

	I have benefited from recent discussions on this general
subject with Anthony Aguirre, Nick Bostrom, Raphael Bousso, Sean
Carroll, David Coule, William Lane Craig, George Ellis, Gary Gibbons,
Steve Giddings, Richard Gott, Jim Hartle, Tomas Kopf, Pavel Krtous,
John Leslie, Andrei Linde, Don Marolf, Joe Polchinski, Mark Srednicki,
Glenn Starkman, Lenny Susskind, Neil Turok, Bill Unruh, Alex Vilenkin,
and others, though I can't remember precisely which of these
discussions were directly relevant to the present paper.  Andrei Linde
kindly made detailed comments on preliminary versions of this
manuscript that made a serious error in the interpretation of his
work.  Financial support was provided by the Natural Sciences and
Engineering Research Council of Canada.

%\newpage


\baselineskip 5pt

\begin{thebibliography}{99}

\bibitem{SQM}  D. N. Page, ``Sensible Quantum Mechanics:  Are Only
Perceptions Probabilistic?'' quant-ph/9506010; Int.\ J.\ Mod.\
Phys.\ {\bf D5}, 583 (1996), gr-qc/9507024.

\bibitem{MS} D.~N.~Page, in {\em Consciousness: New Philosophical
Perspectives}, edited by Q.~Smith and A.~Jokic (Oxford, Oxford
University Press, 2003), pp.\ 468-506, quant-ph/0108039.

\bibitem{Page-in-Carr} D.~N.~Page, in {\em Universe or Multiverse?},
edited by B.~J.~Carr (Cambridge University Press, Cambridge, 2007),
pp.\ 401-419; hep-th/0610101.

\bibitem{Let1} A.~D.~Linde, in {\it  The Very Early Universe}, ed.\
G.~W.~Gibbons, S.~W.~Hawking and S.~Siklos, Cambridge University Press
(1983), pp. 205-249, http://www.stanford.edu/$\sim$alinde/1983.pdf.

\bibitem{Vet1} A.~Vilenkin, Phys.\ Rev.\ {\bf D27}, 2848-2855
(1983).

\bibitem{Set} A.~A.~Starobinsky, in {\em Field Theory, Quantum
Gravity, and Strings}, edited by H.~J.~de Vega and N.~Sanchez,
Lecture Notes in Physics Vol.\ 246 (Springer, Heidelberg, 1986).

\bibitem{Let2} A.~D.~Linde, Mod.\ Phys.\ Lett.\ {\bf A1}, 81-85 (1986).

\bibitem{Let3} A.~D.~Linde, Phys.\ Lett.\ {\bf B175}, 395-400 (1986).

\bibitem{Let4} A.~D.~Linde, Phys.\ Scripta {\bf T15}, 169 (1987).

\bibitem{Let5} A.~D.~Linde, Phys.\ Lett.\ {\bf B249}, 18-26
(1990).

\bibitem{Let6} A.~D.~Linde,  {\it  Particle  Physics  and Inflationary
Cosmology} (Harwood, Chur, Switzerland, 1990), hep-th/0503203.

\bibitem{Let7} A.~D.~Linde,  Sci.\ Am.\ {\bf 271}, 32-39 (1994).

\bibitem{Vet2} A.~Vilenkin, Nucl.\ Phys.\ Proc.\ Suppl.\ {\bf 88},
67-74 (2000), gr-qc/9911087.

\bibitem{Get} A.~H.~Guth, Phys.\ Rept.\ {\bf 333}, 555-574
(2000), astro-ph/0002156.

\bibitem{Vet3} A.~Vilenkin, gr-qc/0409055; {\em Many Worlds in One:
The Search for Other Universes} (Hill and Wang, New York, 2006).

\bibitem{LM} A.~D.~Linde and A.~Mezhlumian, \ Phys.\ Lett.\ {\bf
B307}, 25-33 (1993); Phys.\ Rev.\ {\bf D53}, 4267-4274 (1996),
gr-qc/9511058.

\bibitem{LLM} A.~D.~Linde, D.~A.~Linde, and A.~Mezhlumian, Phys.\
Rev.\ {\bf D49}, 1783-1826 (1994), gr-qc/9306035; Phys.\ Lett.\
{B345}, 203-210 (1995), hep-th/9411111.

\bibitem{GLL} J.~Garcia-Bellido, A.~D.~Linde and D.~A.~Linde, Phys.\
Rev.\ {\bf D50}, 730-750 (1994), astro-ph/9312039.

\bibitem{GL} J.~Garcia-Bellido and A.~D.~Linde, Phys.\ Rev.\ {\bf
D51}, 429-443 (1995),\\hep-th/9408023; Phys.\ Rev.\ {\bf D52},
6730-6738 (1995), gr-qc/9504022.

\bibitem{Vil} A.~Vilenkin, Phys.\ Rev.\ Lett.\  {\bf 74}, 846-849
(1995), gr-qc/9406010; Phys.\ Rev.\ {\bf D52}, 3365-3374 (1995),
gr-qc/9505031; gr-qc/9507018; Phys.\ Rev.\ Lett.\ {\bf 81}, 5501-5504
(1998), hep-th/9806185; hep-th/0602264.

\bibitem{WV} S.~Winitzki and A.~Vilenkin, Phys.\ Rev.\ {\bf D53},
4298-4310 (1996), gr-qc/9510054.

\bibitem{VVW} V.~Vanchurin, A.~Vilenkin, and S.~Winitzki, Phys.\
Rev.\ {\bf D61}, 083507 (2000), gr-qc/9905097.

\bibitem{GV} J.~Garriga and A.~Vilenkin, Phys.\ Rev.\ {\bf D64}, 023507
(2001), gr-qc/0102090; Prog.\ Theor.\ Phys.\ Suppl.\  {\bf 163},
245-257 (2006), hep-th/0508005.

\bibitem{GSVW} J.~Garriga, D.~Schwartz-Perlov, A.~Vilenkin, and
S.~Winitzki, J.\ Cosmol.\ Astropart.\ Phys.\ {\bf 0601}, 017 (2006),
hep-th/0509184.

\bibitem{ELM} R.~Easther, E.~A.~Lim, and M.~R.~Martin, J.\ Cosmol.\
Astropart.\ Phys.\ {\bf 0603}, 016 (2006), astro-ph/0511233.

\bibitem{Teg} M.~Tegmark, J.\ Cosmol.\ Astropart.\ Phys.\ {\bf 0504},
001 (2005), astro-ph/0410281.

\bibitem{FHW} B.~Feldstein, L.~J.~Hall and T.~Watari, Phys.\ Rev.\ {\bf
D72}, 123506 (2005), hep-th/0506235.

\bibitem{Lyth} D.~H.~Lyth, J.\ Cosmol.\ Astropart.\ Phys.\ {\bf 0511},
  006 (2005), astro-ph/0510443.

\bibitem{LMuk} A.~Linde and V.~Mukhanov, J.\ Cosmol.\ Astropart.\ Phys.\
{\bf 0604}, 009 (2006), astro-ph/0511736.

\bibitem{HWY} L.~J.~Hall, T.~Watari and T.~T.~Yanagida, Phys.\ Rev.\
{\bf D73}, 103502 (2006), hep-th/0601028.

\bibitem{Bousso} R.~Bousso, ``Holographic Probabilities in Eternal
Inflation,'' hep-th/0605263.

\bibitem{DKS} L.~Dyson, M.~Kleban, and L.~Susskind, J.\ High Energy
Phys.\ {\bf 0210}, 011 (2002), hep-th/0208013.

\bibitem{GKS} N.~Goheer, M.~Kleban, and L.~Susskind, J.\ High Energy
Phys.\ {\bf 0307}, 056 (2003), hep-th/0212209.

\bibitem{Sus03} L.~Susskind, in {\em Universe or Multiverse?}, edited
by B.~J.~Carr (Cambridge University Press, Cambridge, 2007), pp.
241-260, hep-th/0302219.

\bibitem{lifetime} D.~N.~Page, J.\ Korean Phys.\ Soc.\ {\bf 49},
711-714 (2006), hep-th/0510003.

\bibitem{decay} D.~N.~Page, ``Is Our Universe Likely to Decay within 20
Billion Years?'' \\hep-th/0610079.

\bibitem{HH-challenge} D.~N.~Page, ``Susskind's Challenge to the
Hartle-Hawking No-Boundary Proposal and Possible Resolutions,''
hep-th/0610199.

\bibitem{BBS} L.~Blum, M.~Blum, and M.~Shub, SIAM J.\ Computing, {\bf
15}, 364-383 (1986).

\bibitem{BF} R.~Bousso and B.~Freivogel, ``A Paradox in the Global
Description of the Multiverse,'' hep-th/0610132.

\bibitem{local} R.~Bousso, J.\ High Energy Phys.\ {\bf 11}, 038 (2000),
hep-th/0010252.

\bibitem{LindeBB} A.~Linde, ``Sinks in the Landscape and the
Invasion of Boltzmann Brains,'' hep-th/0611043.

\bibitem{42} D.~Adams, {\it The Hitchhiker's Guide to the Galaxy} (Pan
Books, London, 1979).

\end{thebibliography}
\end{document}